\documentstyle[12pt,twoside]{article}
\advance\hoffset by -7mm
\makeatletter
\@addtoreset{equation}{section}
\makeatother

\renewcommand{\title}[1]{\null\vspace{25mm}

\noindent{\Large{\bf #1}}\vspace{10mm}

\noindent {\large By }}
\newcommand{\authors}[1]{\noindent{\large #1}\vspace{3mm}

}
\newcommand{\address}[1]{\noindent #1\vspace{5mm}

}
\renewcommand{\abstract}[1]{\vspace{19mm}

\noindent{\small{\em Abstract.} #1}\vspace{2mm}

}
\def\bfgr#1{\mbox{{\boldmath $#1$}}}
\def\bfs{\bfgr{\sigma}}
\def\d{{\rm d}}
\def\x{{\bf x}}
\def\p{\partial}
\def\pp{{\overline\partial}}
\def\be{\begin{equation}}
\def\ee{\end{equation}}
\def\ba#1{\begin{array}{#1}}
\def\ea{\end{array}}
\def\bea{\begin{eqnarray}}
\def\eea{\end{eqnarray}}
\def\dfrac#1#2{{\displaystyle\frac{#1}{#2}}}
\def\I{{\cal I}}
\def\J{{\cal J}}
\def\oI{{\overline {\cal I}}}
\def\oJ{{\overline {\cal J}}}
\def\div{{\rm div}}
\def\rot{{\rm rot}}
\def\Div{{\rm Div}}
\def\Rot{{\rm Rot}}
\def\E{{\bf E}}
\def\B{{\bf B}}
\def\D{{\bf D}}
\def\H{{\bf H}}
\def\L{{\cal L}}
\def\oL{{\overline {\cal L}}}
\def\Om#1{\stackrel{#1}{\Omega}}
\def\V#1#2{{\stackrel{#1}{V}}{{}^#2}}
\def\Vd#1#2{{\stackrel{#1}{V}}{{}_#2}}
\def\bfV#1{ \stackrel{#1}{ {\bf V} } }
\def\bV2#1{{\stackrel{#1}{{\bf V}}}{{}^2}}
\def\a#1#2{{\stackrel{#1}{a}}{{}^#2}}
\def\bfa#1{{\stackrel{#1}{{\bf a}}}}
\def\sL#1#2#3{{\stackrel{#1}{L}}{{}^#2_{#3}}}
\def\jj#1#2{{\stackrel{#1}{J}}{{}^#2}}
\def\bfjj#1{{\stackrel{#1}{{\bf J}}}}
\def\A#1#2{\stackrel{\mid #1}{A}_{#2}}
\def\Au#1#2{{\stackrel{\mid #1}{A}}{}^{#2}}
\def\solA#1#2{\stackrel{#1)}{A}_{#2}}
\def\Ssolfi#1{\stackrel{#1}{\phi}}
\def\SsolF#1#2{\stackrel{#1}{F}_{#2}}
\def\SsolE#1#2{\stackrel{#1}{E}_{#2}}
\def\Ssoly#1#2{\stackrel{#1}{y}{{}^{#2}}}
\def\Ssolyd#1#2{\stackrel{#1}{y}{}_#2}

\def\P#1{\stackrel{#1}{P}}

\def\tn#1{\stackrel{#1}{\tau}}
\def\en#1{\stackrel{#1}{e}}
\def\m#1{\stackrel{#1}{m}}
\def\const{{\rm const}}
\def\sts#1{\stackrel{#1}{\sigma}}
\def\T{{\overline T}}
\def\fP{{\overline P}}
\def\rn#1{\stackrel{#1}{r}}
\begin{document}
\mathsurround=2pt

\makeatletter
\global\@specialpagefalse
\def\@oddhead{%
\ifnum\c@page=1
\raisebox{0pt}[\headheight][0pt]{%
\vbox{\hbox to\textwidth{\strut
To appear in Helv. Phys. Acta \hfill  http://xxx.lanl.gov/ps/hep-th/9705075}
\hrule}}%
\fi
\ifnum\c@page>1
\raisebox{0pt}[\headheight][0pt]{%
\vbox{\hbox to\textwidth{\strut A.A. Chernitskii
\hfill  http://xxx.lanl.gov/ps/hep-th/9705075}
\hrule}}%
\fi
}
\def\@evenhead{\raisebox{0pt}[\headheight][0pt]{%
\vbox{\hbox to\textwidth{\strut  To appear in Helv. Phys. Acta
\hfill A.A. Chernitskii}
\hrule}}}

\def\@oddfoot{\reset@font\rm\hfill \thepage}
\def\@evenfoot{\reset@font\rm \thepage\hfill}
\makeatother

\title{Nonlinear Electrodynamics with Singularities\\[2mm]
(Modernized Born-Infeld Electrodynamics)}
\authors{Alexander A. Chernitskii \footnote{E-mail: aa@cher.etu.spb.ru}}
\address{Department of Physical Electronics,
St.Petersburg Electrotechnical University,\\
Prof. Popov str. 5, St.Petersburg 197376,  Russia }

\abstract{
Born-Infeld nonlinear electrodynamics are considered.
Main attention is given to existence of singular point at static
field configuration that M.Born and L.Infeld are considered as a model of
electron. It is shown that such singularities are forbidden
within the framework of the Born-Infeld model. It is proposed a
modernized action that make possible an existence of the singularities.
It is obtained main relations in view of the singularities.
In initial approximation this model gives the usual linear
electrodynamics with point charged particles.
}

\section{Introduction}
In the article by M.Born and L.Infeld \cite{BornInfeld} a model for nonlinear
electrodynamics was considered. This model is named now as
Born-Infeld electrodynamics. The article \cite{BornInfeld} contain
some arguments about the necessity for nonlinear generalization of
electrodynamics and possible connection between the such theories and
quantum mechanics.
At the present time the interest to similar models continues.
In this connection it should be noted a series of articles
by A.O.~Barut about connection between classical particle-like field
configurations, real particles and quantum theory.
For example, see the articles
\cite{Barut:homega,Barut:withouth} by A.O.~Barut and the article
\cite{Barut:PartLike} by A.O.~Barut and A.J.~Bracken.
In this cited articles the authors consider a linear theory of field but the
stated ideas may be used also for nonlinear models. Moreover, nonlinearity
create interactions of various types between particle-like solutions
when we use a perturbation method. Thus there is a possibility of
description of real interaction of particles with help of the such
models.
\par
The Born-Infeld model has some attractive properties. In particular,
a field configuration that appropriate to the
electron has a finite energy in framework of Born-Infeld electrodynamics,
in contrast to the case of usual linear electrodynamics.
But the components of electromagnetic field of this configuration
has a discontinuity at a point for Born-Infeld model,
the four-vector potential being everywhere continuous. However,
the field configuration at the singular point is not solution
to the field equations which are in the article  \cite{BornInfeld}.
In the present work we derive conditions for stationarity of the
Born-Infeld action in the case when the field may have singularities.
It appears, that it is necessary to modernize
the Born-Infeld action in order that the model allowed the existence of such
singularities and, hence, could be suitable for the description of real
particles. We propose a modernized action and we obtain the main relations
for particles-singularities.

\section{Born-Infeld Electrodynamics}
Let us state the main relations about the Born-Infeld standard model
and introduce some notations which we shall use.
The Born-Infeld action has the following form \cite{BornInfeld,TonnelatMA}.
\bea
S_{BI} {}={} \int \left[\,\sqrt{|\det(g_{\mu\nu}  {}+{}  \alpha\,F_{\mu\nu})|}  {}-{}
\sqrt{|\det(g_{\mu\nu})|}\,\right]\,(\d x)^4  {}={}
\int \left( \L  {}-{}  1 \right)\,\sqrt{|g|}\;(\d x)^4
\label{ActionBI}
\eea
\bea
\mbox{where}
& & F_{\mu\nu}  {}\equiv{}  \frac{\p A_\nu}{\p x^\mu}   {}-{}
\frac{\p A_\mu}{\p x^\nu}
\quad\mbox{ is electromagnetic field tensor components}\nonumber\\[2.5mm]
& & g_{\mu\nu}  \quad\mbox{ is metric tensor components; }
\mbox{ the Greec indexes take value } 0,\,1,\,2,\,3\nonumber\\[2.5mm]
& & \L {}\equiv{}   \sqrt{|\,1 {}-{}  \alpha^2\,\I  {}-{}  \alpha^4\,\J^2\,|}\quad;\quad
\I {}\equiv{}  \frac{1}{2}\,F_{\mu\nu}\,F^{\nu\mu}
\quad;\quad
\J {}\equiv{}  \sqrt{\frac{F}{|g|}}  {}={}
\frac{1}{8}\,\varepsilon_{\mu\nu\sigma\rho}\, F^{\mu\nu} F^{\sigma\rho}
\nonumber\\[2.5mm]
& &
F  {}\equiv{}  \det (F_{\mu\nu})\quad;\qquad g  {}\equiv{}  \det (g_{\mu\nu})\quad;\qquad
(\d x)^4  {}\equiv{}  \d x^0 \d x^1 \d x^2 \d x^3 \nonumber\\[2.5mm]
& &
\!\!
\left.
\ba{rcl}
\varepsilon_{\mu\nu\sigma\rho}  & \equiv & \pm\sqrt{|g|}
\nonumber\\[5pt]
\varepsilon^{\mu\nu\sigma\rho}  & = & \mp
\dfrac{1}{\sqrt{|g|}}
\ea
\right\}
\ba{l}
\mbox{ here there is the }
{}^{\mbox{\phantom{bo}top}}_{\mbox{bottom}}\mbox{ sign, if }
\mu\nu\sigma\rho
\mbox{ is }{}^{\mbox{even}}_{\mbox{odd}}\\[4mm]
\mbox{ permutation of indexes } 0123
\ea
\nonumber
\eea
A condition for stationarity of the action
(\ref{ActionBI}) is the following Eulerian system of equations.
\be
\frac{\p}{\p x^\mu}\,\sqrt{|g|}\; f^{\mu\nu} {}={} 0
\label{EulerBI}
\ee
where
\be
f^{\mu\nu} {}\equiv{}  \frac{1}{\alpha^2}\,\frac{\p\L}{\p(\p_\mu A_\nu)} {}={}
\frac{1}{\L}\,\left(F^{\mu\nu}  {}-{}
\frac{\alpha^2}{2}\,\J\,\varepsilon^{\mu\nu\sigma\rho}\,F_{\sigma\rho}\right)
\label{Def:f}
\ee
The expression for the tensor components $F_{\mu\nu}$ as
functions of the tensor components $f_{\mu\nu} $ has the following form.
\bea
F_{\mu\nu} {}={}
\frac{1}{\oL}\,\left(f_{\mu\nu}  {}+{}
\frac{\alpha^2}{2}\,\oJ\,\varepsilon_{\mu\nu\sigma\rho}\,f^{\sigma\rho}\right)
\eea
where
\bea
\left\{
\ba{ccl}
\oL & \equiv & \sqrt{|1 {}-{} \alpha^2 \,\oI {}-{} \alpha^4 \,\oJ^2 |}\\[11pt]
\oI & \equiv & \dfrac{1}{2}\,f^{\mu\nu}\,f_{\mu\nu}\\[11pt]
\oJ & \equiv &
\dfrac{1}{8}\,\varepsilon_{\mu\nu\sigma\rho}\,f^{\mu\nu}\,f^{\sigma\rho}
\ea
\right.
\quad\Longrightarrow\quad
\left\{
\ba{ccl}
\oJ &   {}={}  &  \J\\[11pt]
\oL\,\L &  = &  1 {}+{}  \alpha^4 \, \J^2
\ea
\right.
\eea
In view of the electromagnetic tensor definition we have also
the following relation.
\be
\varepsilon^{\mu\nu\sigma\rho}\,\frac{\p F_{\sigma\rho}}{\p x^\nu}  {}={}  0
\label{Eq:F}
\ee
If we introduce the notations for space vectors of the field\\
(the latin indexes take values $1,\,2,\,3$)
\be
\left\{
\ba{rcl}
E_i&\equiv& F_{i0}\\[2.5mm]
B^i&\equiv&  -\dfrac{1}{2}\, \varepsilon^{0ijk}\, F_{jk}\\[5mm]
F_{ij} &=& \varepsilon_{0ijk}\, B^k
\ea
\right.
\qquad;\qquad
\left\{
\ba{rcl}
D^i&\equiv&  f^{0i}\\[2.5mm]
H_i&\equiv&
\dfrac{1}{2}\,\varepsilon_{0ijk}\, f^{jk}\\[5mm]
f^{ij} &=& -\varepsilon^{0ijk}\, H_k
\ea
\right.
\label{Def:EBDH}
\ee
then the system (\ref{EulerBI}), (\ref{Eq:F}) may be written in
the following form.
\be
\left\{
\ba{rcl}
\Div \B  &=& 0\\[2.5mm]
\phantom{-}\pp_0 \,\B  {}+{}  \Rot\E  &=& 0\\[2.5mm]
\Div \D  &=& 0\\[2.5mm]
-\pp_0 \,\D  {}+{}  \Rot\H  &=& 0
\ea
\right.\quad\qquad
\mbox{where}\qquad
\left\{
\ba{rcl}
\pp_\mu & \equiv &
\dfrac{1}{\sqrt{|g|}}\,\dfrac{\p}{\p x^\mu}\,\sqrt{|g|}\\[8mm]
\Div \B &\equiv& \pp_i\, B^i \\[3.5mm]
(\Rot \E)^i &\equiv& -\varepsilon^{0ijk}\,\p_j E_k\\[2.5mm]
\p_\mu  & \equiv & \dfrac{\p}{\p x^\mu}
\ea
\right.
\label{Eq:Maxwell}
\ee
Here the definitions $\Div$ and $\Rot$ (\ref {Eq:Maxwell}) include
$\sqrt{|g|}$ for the space-time in contrast to the usual definitions
$\div$ and $\rot$ for the space. It is evident that these definitions
coincide for the case with $|g_{00}|  {}={}  1$ and $g_{0i}  {}={} 0$.
If in addition we have $g {}\neq{} g(x^0)$ then
the system of equations (\ref{Eq:Maxwell}) has the form of the usual
Maxwell's system of equations.  It should be noted that it is for the
definitions (\ref{Def:EBDH}) that we have the form of equations system
(\ref{Eq:Maxwell}), the distinction of the top and bottom indexes
being essential. The forms of the definitions (\ref{Def:EBDH}) and
equations system (\ref{Eq:Maxwell}) does not depend on the form of
dependence between tensor components $f^{\mu\nu}$ and $F^{\sigma\rho}$
(\ref{Def:f}). Thus the foregoing is right also for usual linear
electrodynamics that we have through linearization the relations
(\ref{Def:f}): $f^{\mu\nu} {}={} F^{\mu\nu}$. The system of equations
(\ref{Eq:Maxwell}) is right for any space-time metric.
\par
According to formulas (\ref{Def:f}), (\ref{Def:EBDH})
the components of the vectors $\D$, $\H$ and $\E$, $\B$ are interconnected
by means of the following formulas.
\be
\left\{
\ba{ccl}
D^i & = & \dfrac{1}{\L}\,
\left[\vphantom{\dfrac{1}{2}}\left(
\vphantom{\frac{1}{2}} g^{0j}\,g^{i0}  {}-{}
g^{00}\,g^{ij}\right)\,E_j  {}+{}
g^{0j}\,g^{ip}\,\varepsilon_{0jpk}\, B^k  {}+{}
\alpha^2\,\J\, B^i \right]\\[4mm]
H_i & = & \dfrac{1}{\L}\,
\left[\dfrac{1}{2}\,
\varepsilon_{0ijk}\,g^{jl}\,g^{kp}\,\varepsilon_{0lpm}\, B^m  {}+{}
\dfrac{1}{2}\,\varepsilon_{0ijk}\,\left(g^{jl}\,g^{k0}  {}-{}
g^{j0}\,g^{kl}\right)\,E_l  {}+{}
\alpha^2\,\J\, E_i \right]
\ea
\right.
\label{GenMatEq}
\ee
If we substitute $\alpha {}={} 0$, $\L {}={} 1$ in (\ref{GenMatEq}) we have the
appropriate expressions for the case of the linear electrodynamics.
\par
The electromagnetic invariant are expressed through
the field space vectors components with the following formulas.
\bea
\ba{ccl}
\I & = & \left(g^{k0}\,g^{0l}  {}-{}  g^{kl}\,g^{00}\right)\,E_k\,E_l  {}+{}
\dfrac{1}{2}\,\varepsilon_{0ijk}\,g^{jm}\,g^{in}\,\varepsilon_{0mnl}\,B^k\,B^l
{}-{}\\[9pt]
& & \qquad \qquad\qquad\qquad\qquad{}-{} \varepsilon_{0ijk}\,\left[
\dfrac{1}{2}\,
\left(g^{jl}\,g^{i0}  {}-{}  g^{j0}\,g^{il}\right)  {}-{}  g^{li}\,g^{0j}
\right]\,E_l\,B^k\\
\J & = & E_i\,B^i
\ea
\eea
If $|g^{00}| {}={} 1$ and $g^{0j} {}={} 0$
we have the simple relations interconnecting the
vectors $\D$, $\H$ and $\E$, $\B$.
\bea
\left\{
\ba{ccl}
\D &=& \dfrac{1}{\L}\,(\E  {}+{}  \alpha^2\,\J \B)\\[9pt]
\H &=& \dfrac{1}{\L}\,(\B  {}-{}  \alpha^2\,\J \E)
\ea
\right.\quad;\qquad
\left\{
\ba{ccl}
\E &=& \dfrac{1}{\oL}\,(\D  {}-{}  \alpha^2\,\J \H)\\[9pt]
\B &=& \dfrac{1}{\oL}\,(\H  {}+{}  \alpha^2\,\J \D)
\ea
\right.
\label{MatEq}
\eea
In this case the expressions for the electromagnetic invariants have
the following form.
\bea
\left\{
\ba{ccl}
\I &=& \E\cdot\E  {}-{}  \B\cdot\B\\[9pt]
\J &=& \E\cdot\B
\ea
\right.\quad;\qquad
\left\{
\ba{ccl}
\oI &=& \H\cdot\H  {}-{}  \D\cdot\D\\[9pt]
\oJ &=& \H\cdot\D
\ea
\right.
\eea
(Expressions for $\J$ and $\oJ$ in the case of general metric are the same).
\par
A little more about the linear electrodynamics. If we take
a variational principle as the basis of the model, then the usual relations
for the linear electrodynamics $\D {}={} \E$, $\H {}={} \B$ are right provided we have
$|g_{00}|  {}={}  1$ and $g_{\mu\nu} {}={} 0$ for $\mu\neq\nu$.
In the case of general metric we must take the relations (\ref{GenMatEq})
(with $\alpha {}={} 0$, $\L {}={} 1$ in this case).
\par
In the article \cite{BornInfeld} it is proposed the following
spherically symmetrical field configuration as a solution to the model
that appropriate to electron.
\be
E_r  {}={}  \frac{e}{\sqrt{\alpha^2 \, e^2  {}+{}  r^4}}\qquad;\qquad
D_r  {}={}  \frac{e}{r^2}
\label{Sol:Stat}
\ee
where $r$ is the radial coordinate,
and the index $r$ denote the radial component of vector.
\par
As we see, the vector $\E$ has discontinuity in the beginning of coordinates.
But the function $A_0(x)$ is continuous everywhere and the action
is finite.
It is obvious that (\ref{Sol:Stat}) is a solution to the system of equations
(\ref{Eq:Maxwell}) everywhere except the coordinate center.
In the beginning of coordinates $\div\D$ is infinity for the field
configuration (\ref{Sol:Stat}).
\par
With Lorentz transformations we have the appropriate to (\ref{Sol:Stat})
moved field configuration, that has the following form in a cartesian
system of coordinates.
\be
F_{\mu\nu} {}={} (L^{i}_{.\mu} L^{0}_{.\nu} {}-{} L^{0}_{.\mu} L^{i}_{.\nu})\, E_i(r)
\quad;\quad r {}={} \sqrt{y^i y_i}
\quad;\quad y^i {}={} L^i_{.j} \left(x^j  {}-{}  V^j x^0\right)
\label{Sol:Move}
\ee
Here
$E_i$ is the cartesian components appropriate to (\ref{Sol:Stat}),
$\{y^i\}$ is own cartesian coordinate system of the singularity
moving with it together,
$V^j$ is components of velocity of the singularity,
$||L_{\mu\nu}||$ is matrix of Lorentz transformations appropriate to boost:
\bea
L^0_{.0}  {}={} \frac{1}{\sqrt{1 {}-{} \bV2{}}}\;\;;\;\;\;
L^0_{.i}  {}={}  L_{i0}  {}={}
 -\frac{\Vd{}{i}}{\sqrt{1 {}-{} \bV2{}}}\;\;;\;\;\;
L^i_{.j} {}={} \delta^i_j {}+{} \left(\frac{1}{\sqrt{1 {}-{} \bV2{}}}  {}-{}  1\right)\,
\frac{V^i V_j}{{\bf V}^2}\;\;\;
\label{Def:Lorenz}
\eea
We use a metric with the signature $\{-+++\}$.
\par
Because $\J {}={} 0$ for the field configuration (\ref{Sol:Move}), it
 satisfies (out of the singularity) also
a more simple system of equation then (\ref{Eq:Maxwell}),
(\ref{GenMatEq}) or (\ref{Eq:Maxwell}), (\ref{MatEq}). We obtain this system
if substitute $\J {}={} 0$ into relations (\ref{GenMatEq}), (\ref{MatEq}) and
take $\L {}={} \sqrt{|\,1 {}-{}  \alpha^2\,\I\,|}$, $\oL {}={} \sqrt{|\,1 {}-{}  \alpha^2\,\oI\,|}$.
An action has the following form \cite{BornInfeld} for this case.
\bea
S^{'}_{BI} {}={} \int \left[\,\sqrt{|\det(g_{\mu\nu}  {}+{}  \alpha\,F_{\mu\nu})  {}-{}
\det(\alpha\,F_{\mu\nu})|}  {}-{}
\sqrt{|\det(g_{\mu\nu})|}\,\right]\,(\d x)^4
\label{ActionBIs}
\eea

\section{Conditions for Stationarity of the Action\\
with Singular Points of Field}
Let the functions $A_\mu(x^\nu)$ that give stationarity
to the action functional (\ref{ActionBI}) or (\ref{ActionBIs})
be continuous and have $N$
singular points as discontinuity of derivatives on the coordinates.
We assume that the singularities exist for all time.
We do not consider here a possible process of birth-destruction of
the singularities. Thus in any moment of time each
$n$-th singularity has certain space coordinates. We shall use the notation
$\a{n}{i}(x^0)$ for it. Let us enclose each singularity in a small
closed surface moving with the singularity together.
Subsequently we shall contract these surfaces to the points.
Let us consider the following action.
\be
{\overline S}_{BI}  {}={}  \int \limits_{-\infty}^{+\infty} \d x^0
\int\limits_{\overline\Omega} (\L {}-{} 1)\,\sqrt{|g|}\;(\d x)^3
\label{bAction}
\ee
where $(\d x)^3  {}\equiv{}  \d x^1 \d x^2 \d x^3$,
${\overline\Omega} $ is the all three-dimensional space
with the excluded regions bounded by the surfaces which enclose the
singular points.
\par
We shall make a variation of the functional (\ref{bAction}) through
variations of the field functions $A_\mu(x)$ and
trajectories of the singularities. Here we have a moving boundary
variational problem with free variation on the boundary \cite{CurantHilbert}.
As it is usual, we shall replace
$A_\mu(x) \to A_\mu (x)  {}+{}  \varepsilon\, \delta A_\mu (x)$ and
$\a{n}{i} (x^0) \to \a{n}{i} (x^0)  {}+{}  \varepsilon\, \delta \a{n}{i} (x^0) $
in the action (\ref{bAction}), considering that $\delta A_\mu (x)  {}={} 0$
and $\delta \a{n}{i} (x^0)  {}={}  0$ at infinity.
We differentiate the action on
$\varepsilon$ and take $\varepsilon {}={} 0.$ Then we make the partial
integration and obtain the following variation of the action (\ref{bAction}).
\bea
\delta {\overline S_{BI}} & {}={} &\int \limits_{-\infty}^{+\infty} \d x^0
\left\{\,
-\alpha^2\,\int\limits_{\overline\Omega} \frac{\p}{\p x^\mu}\,
\left(\sqrt{|g|}\, f^{\mu\nu}\right)\,
\delta A_\nu\,(\d x)^3
\right.{}-{}\label{bActionA}\\
& &\quad
\left. {}-{}\sum\limits_{n {}={} 1}^N\,\left[
\alpha^2\,\left(\int\limits_{\sts{n}}
f^{i\nu}\,\delta A_\nu\,\d\sts{n}_i {}-{}
\int\limits_{\sts{n}}
f^{0\nu}\,\delta A_\nu\,\V{n}{i}\,\d\sts{n}_i
\right) {}+{}
\int\limits_{\sts{n}}
(\L {}-{} 1)\,\delta \a{n}{i}\;\d\sts{n}_i\right]
\,\right\}               \nonumber
\eea
\bea
\mbox{Here}
& &\sts{n} \mbox{ \quad is \quad  } \mbox{closed surface enclosing
the $n$-th singularity, it make the integra-}\nonumber\\
& & \phantom{\sts{n} \mbox{ \quad is \quad }}
\mbox{tion on external (relative to the singular point)
side of this surface}\nonumber\\[4pt]
& &
\!\!
\left.
\ba{ccl}
\d {{\sts{n}}}_1 &=&\pm\sqrt{|g|}\,\d x^2 \d x^3\\
\d {{\sts{n}}}_2 &=&\pm\sqrt{|g|}\,\d x^1 \d x^3\\
\d {{\sts{n}}}_3 &=&\pm\sqrt{|g|}\,\d x^1 \d x^2
\ea
\right\}
\ba{ccl}
& &\mbox{where it take the sign "$+$", if we have an}\\
& &\mbox{acute angle between $i$-th coordinate axis}\\
& &\mbox{and external normal to the surface $\sts{n}$ and}\\
& &\mbox{it take sign "$-$", if the angle is obtuse}
\ea\\[12pt]
& & \mbox{with the differential form concept the
following expression can be written}\nonumber\\[4pt]
& &\d {{\sts{n}}}_i  {}={} \dfrac{1}{2}\,\varepsilon_{0ijk}\, \d x^j \wedge \d x^k\\
& & \V{n}{i}  {}\equiv{}  \frac{\d \a{n}{i}}{\d x^0}
\eea
We have the last integral on the surfaces in (\ref{bActionA}) with
differentiation of location of the integration region border
 by $\varepsilon$ \cite{CurantHilbert}.
\par
When we contract the all surfaces $\sts{n}$ to the points
we have ${\overline \Omega} \rightarrow {\dot\Omega}$, where
${\dot\Omega} $ is the whole three-dimensional space without the
singular points.
Replacement of $\Omega$ in ${\dot\Omega}$ is similar to partition of
argument value interval of some function having discontinuity of
derivative on two intervals: at the right and at the left from the
singularity. Exception of the singular points means only that
the infinite values of derivatives of the functions $f^{\mu\nu} (x)$
are not included in the integral. Thus it is evident that if
${\overline\Omega} \rightarrow {\dot\Omega}$ then
${\overline S}_{BI} \rightarrow S$ and
$\delta {\overline S}_{BI} \rightarrow \delta S_{BI}$.
\vskip 5pt
\par
The functions $\delta A_\nu (x)$ are continuous everywhere, hence,
we can factor out these functions from the integral on the surfaces
$\sts{n}$ when we contract these surfaces to the points.
Let us introduce the following designation for charge and
current of the singularity.
\bea
\en{n} {}\equiv{}  \frac{1}{4 \pi}\,\lim_{{\sts{n}} \to 0}\,
\int\limits_{{\sts{n}}}
 f^{0 i}\,\d{{\sts{n}}}_i
\quad;\qquad
\jj{n}{l}  {}\equiv{}  \frac{1}{4 \pi}\, \lim_{{\sts{n}} \to 0}\,
\int\limits_{{\sts{n}}}
\left( f^{li}\,\d{{\sts{n}}}_i {}-{}
 f^{l0}\,\V{n}{i}\,\d{{\sts{n}}}_i \right)
\eea
According to the notations (\ref{Def:EBDH}) we have the following
expressions.
\bea
\en{n} {}\equiv{} \frac{1}{4 \pi}\,\lim_{{\sts{n}} \to 0}\, \int\limits_{{\sts{n}}}
\D\cdot \d{{\stackrel{n}{\bfs}}} \quad;\qquad
\bfjj{n} {}\equiv{} \frac{1}{4 \pi}\,
 \lim_{{\sts{n}} \to 0}\, \int\limits_{{\sts{n}}}
\left[
\D \left(\bfV{n} {}\cdot{} \d{{\stackrel{n}{\bfs}}} \right)
{}-{} \H\times \d{{\stackrel{n}{\bfs}}} \right]
\eea
where \qquad
$(\H\times \d{{\stackrel{n}{\bfs}}})_i {}\equiv{}
\varepsilon_{0ijk}\,H^j\,\d\!\sts{n}{}^k$\quad;\qquad
$(\H\times \d{{\stackrel{n}{\bfs}}})^i {}={} -\varepsilon^{0ijk}\,H_j\,%
\d\!\sts{n}_k$\\
It is clear that if we take a non-singular as $n$-th point then
$\en{n} {}={} 0$, $\bfjj{n} {}={} 0$.
\par
As result we have variation of the action (\ref{ActionBI}) in the
following form.
\bea
\delta S_{BI} {}&=&{} \int \limits_{-\infty}^{+\infty} \d x^0
\left\{-\alpha^2\,\int\limits_{\dot\Omega}
\frac{\p}{\p x^\mu}\left( \sqrt{|g|}\,f^{\mu\nu}\right)\,
\delta A_\nu\,(\d x)^3
\right. {}+{} \label{bActionAA}\\
& &
\left.
{}+{}\sum\limits_{n {}={} 1}^N\,
\left[ \alpha^2\,4 \pi \left(\en{n} \delta A_0 (x^0,\bfa{n}) {}+{}
\jj{n}{i}\, \delta A_i (x^0,\bfa{n})\right) {}-{}
\lim_{{\sts{n}} \to 0}\,\int\limits_{\sts{n}}
\L\;\delta \a{n}{i}\,\d\!\sts{n}_i\right]
\,\right\}
\nonumber
\eea
According to the general principles of calculus of variations
\cite{CurantHilbert} we can take $\delta A_\nu (x^0, \bfa {n}) {}={} 0$
and $\delta \a{n}{i} {}={} 0$ at first.
Thus if $\delta S {}={} 0$ then the second
term in (\ref{bActionA}) is zero. Hence, the field satisfies the
homogeneous system of equations (\ref{EulerBI}) outside of the
singular points. Now we take $\delta \a{n}{i} \neq 0$.
As result we have the following condition for each singularity.
\be
\lim_{{\sts{n}} \to 0}\,\int\limits_{\sts{n}}
\L\;
\d{{\stackrel{n}{\bfs}}} {}={} 0
\label{Cond:limLeq0}
\ee
As we see the field configuration (\ref{Sol:Stat}), (\ref{Sol:Move})
satisfies this condition.
\par
Now if we take $\delta A_\nu (x^0, \bfa{n}) \neq 0$ for any one $n$-th
singularity then we have $\en{n} {}={} \jj{n}{i} {}={} 0$ that
are boundary natural conditions for this variational problem.
It is evident that these conditions forbid existence of
the solution (\ref{Sol:Stat}), (\ref{Sol:Move}) for
which $\en{n} {}\neq{} 0$, $\jj{n}{i} {}\neq{} 0$.
\par
The same conclusion can be obtained if we consider
$\pp_\mu f^{\mu\nu} (x) $ as
distributions or generalized functions. As it is known, a partial
integration is allowable for the class of generalized functions.
Thus if we have the stationary action (\ref{ActionBI})
then the system of equations (\ref{EulerBI}) should
be satisfied everywhere. On the other hand the field
configuration (\ref{Sol:Stat}), (\ref{Sol:Move})
satisfies the system of equations
(\ref{EulerBI}) only outside of the singular point.

\section{Modernized Action, Field Equations\\ and Charge Conservation}

We can modernize the action (\ref{ActionBI}) for the field configuration
(\ref{Sol:Stat}), (\ref{Sol:Move}) could be the solution of model equations.
Let us add to the action (\ref{ActionBI}) the terms that compensate
the terms with charges and currents of the singularities
in the expression (\ref{bActionAA}).
Thus a modernized action has the following form.
\be
S {}={} \int \left( \L  {}-{}  1 \right)\,\sqrt{|g|}\;(\d x)^4
 {}-{} 4 \pi \, \alpha^2\,\sum\limits_{n {}={} 1}^N\,
\int \limits_{-\infty}^{+\infty} \left(
\en{n}\;\A{n}{0} {}+{}
\jj{n}{i}\,\A{n}{i}
\right)\,\d x^0
\label{Action1}
\ee
Here we introduce the following notation.
\be
\A{n}{\nu}  {}\equiv{}  A_\nu(x^0,\bfa{n})
\label{Def:An}
\ee
We can use Dirac $\delta$-function and write the action (\ref{Action1})
in the following form.
\be
S {}={} \int \left( \L  {}-{}  1  {}-{}
4 \pi\,\alpha^2\,A_\mu\,\jmath^{\mu} \right)\,
\sqrt{|g|}\;(\d x)^4
\label{Action2}
\ee
where
\be
\jmath^0 {}\equiv{} \frac{1}{\sqrt{|g|}}\,
\sum\limits_{n {}={} 1}^N\,\en{n}\, \delta(\x {}-{} \bfa{n})\quad;\qquad
\jmath^i {}\equiv{} \frac{1}{\sqrt{|g|}}\,
\sum\limits_{n {}={} 1}^N\,\jj{n}{i}\, \delta(\x {}-{} \bfa{n})
\label{Def:j}
\ee
We use the following definition for three-dimensional $\delta$-function.
\bea
\int\limits_{\Om{n}} f(\x)\,\delta(\x {}-{} \bfa{n})\,(\d x)^3  {}\equiv{}
\lim_{{\sts{n}} \to 0}\,\left\{\frac{1}{|\sts{n}|}\,
\int\limits_{{\sts{n}}}
f(\x)\;\d{{\sts{n}}} \right\} {}\equiv{}
\left\langle\vphantom{\int} f(\x) \right\rangle_n
\quad;\qquad |\sts{n}|  {}\equiv{}  \int\limits_{{\sts{n}}}\d{{\sts{n}}}
\label{Def:delta}
\eea
Here $\Om{n}$ is a region of the space $\Omega$ including the point
$\x  {}={}  \bfa{n}$,
$\d{{\sts {n}}} $ is an area element of the surface $\sts{n}$,
$|\sts{n}|$ is an area of the whole surface $\sts{n}$.
We suppose that function $f(\x)$ may have discontinuity at
the point $\x  {}={}  \bfa{n}$.
\par
First we can consider $\en{n}$, $\jj{n}{i}$ and $\a{n}{i}$
as some given functions of time.
Then making a variation of the field functions in the action (\ref{Action2})
we obtain the following field equation.
\be
\pp_\mu \, f^{\nu\mu} {}={}  4\pi\,\jmath^\nu
\label{EulerMy}
\ee
Then the system of equations for the field vectors $\E,\,\B,\,\D,\,\H$
has the following form.
\bea
\left\{
\ba{rcl}
\Div \B  &=& 0\\
\pp_0 \, \B  {}+{}  \Rot\E  &=& 0
\ea
\right.\quad;\qquad
\left\{
\ba{rcl}
\Div \D  &=& 4\pi\,\jmath^0\\
-\pp_0 \, \D  {}+{}  \Rot\H  &=& 4\pi\,{\bfgr\jmath}
\ea
\right.
\eea
Let us act to the system of equations (\ref{EulerMy})
by the operator $\pp_\nu$ with convolution on the index $\nu$:
$\pp_\nu \pp_\mu\,f^{\nu\mu}  {}={} 4\pi\, \pp_\nu\,\jmath^\nu$.
Here left part is zero because $f^{\mu\nu}  {}={} -f^{\nu\mu}$
and it is possible for distribution to change an order of differentiation
by different coordinates.
Thus we have the following conservation law.
\be
 \pp_\nu\,\jmath^\nu  {}={}  0
\label{ConsLaw:Current0}
\ee
Substituting definition of $\jmath^\nu$ (\ref{Def:j})
into (\ref{ConsLaw:Current0}) we obtain the following relation.
\be
\sum\limits_{n {}={} 1}^N\,\left[\left(\frac{\d\en{n}}{\d x^0}\right)
\delta(\x {}-{} \bfa{n}) {}+{}
\left(\jj{n}{i} {}-{} \en{n}\,\V{n}{i} \right)
\frac{\p}{\p x^i}\,\delta(\x {}-{} \bfa{n})\right] {}={} 0
\label{ConsLaw:Current}
\ee
Let us integrate this relation on a region of the space
 including only one $n$-th singularity in some any moment of time.
As result we obtain that the charge of an individual singularity
is conserved.
\be
\frac{\d\en{n}}{\d x^0}  {}={}  0 \quad \Longrightarrow \quad \en{n} {}={} \const
\label{ConsLaw:Charge}
\ee
\par
Substituting (\ref{ConsLaw:Charge}) into
(\ref{ConsLaw:Current}) we obtain
\be
\bfjj{n}  {}={} \en{n}\,\bfV{n}
\label{Def:Ji}
\ee
Let us rewrite the action (\ref{Action1}) in the following form
using (\ref{ConsLaw:Charge}), (\ref{Def:Ji}).
\be
S {}={} \int \left( \L  {}-{}  1 \right)\,\sqrt{|g|}\;(\d x)^4
{}-{} 4\pi\,\alpha^2\,\sum\limits_{n {}={} 1}^N\, \en{n}
\int \limits_{-\infty}^{+\infty}
\left(\A{n}{0} {}+{}
\frac{\d \a{n}{i}}{\d x^0}\,\A{n}{i}
\right) \d x^0
\label{Action3}
\ee
Introducing an integration on the trajectories of the singularities we can
write the action (\ref{Action3}) in the following form.
\be
S {}={} \int \left( \L  {}-{}  1 \right)\,\sqrt{|g|}\;(\d x)^4
 {}-{} 4\pi\,\alpha^2\,\sum\limits_{n {}={} 1}^N\, \en{n}
\int \limits_{\tn{n}} A_\mu\,\d x^\mu
\label{Action4}
\ee
where $\tn{n}$ is trajectory of $n$-th singularity.

\section{Motion Equations of the Singularities}

Let us make a variation of the action (\ref{Action2}) on the
functions $\a{n}{i}(x^0)$.
We suppose that the condition (\ref{Cond:limLeq0}) and
the equation (\ref{EulerMy}) are satisfied.
Taking into account the definition
(\ref{Def:j}) and relation (\ref{Def:Ji}) we obtain the following condition.
\bea
F_{i0}\;\jmath^0 {}+{} F_{ij}\;\jmath^j {}={}  0
\label{Eq:GMove0}
\eea
Let us integrate this condition on a region of the space
 including only one $n$-th singularity in some any moment of time.
As result we have the following conditions for each singularity.
\be
\left\langle\vphantom{\V{n}{j}} F_{i0} \right\rangle_n {}+{}
\left\langle\vphantom{\V{n}{j}} F_{ij} \right\rangle_n\,\V{n}{j}    {}={} 0
\label{Eq:Move2}
\ee
Here the angle brackets $\langle ... \rangle_n$ is the averaging of
function near the $n$-th singular point (see definition of
$\delta$-function (\ref{Def:delta})).
Remember that the functions $F_{\mu\nu}$ have not single value at
the singular points.
The condition (\ref{Eq:Move2}) may be obtained also without using the
$\delta$-function.
We can substitute the definition of $\delta$-function (\ref{Def:delta})
into action (\ref{Action2}) and derive the condition (\ref{Eq:Move2})
directly.
From (\ref{Eq:Move2}), it is evident that
\be
\V{n}{i} \,\left\langle\vphantom{\V{n}{j}} F_{i0} \right\rangle_n  {}={} 0
\label{Eq:Move2a}
\ee
Using again the definitions of $\delta$-function (\ref{Def:delta})
and current density (\ref{Def:j}), (\ref{Def:Ji}) we can pool
 (\ref{Eq:Move2}) and (\ref{Eq:Move2a}) to the following formula.
\bea
F_{\mu\nu}\,\jmath^\nu  {}={}  0
\label{Eq:GMove}
\eea
Using the definition (\ref{Def:EBDH}), we can write
the condition (\ref{Eq:Move2}) in the following form.
\be
\left\langle\vphantom{\bfV{n}} \E \right\rangle_n {}+{}
\bfV{n} {}\times{} \left\langle\vphantom{\bfV{n}} \B \right\rangle_n  {}={} 0
\label{Eq:Move1}
\ee
\par
We can search a solution with $N$ singularity by
a perturbation method. As initial approximation we shall take
a sum of the solutions with one singularity, the free
parameters of which being dependent from time.
The components of velocity $\V{n}{i}$ are free parameters
for the solution of type (\ref{Sol:Move}).
So, the initial approximation has the following form.
\bea
A_{\mu}(x) {}={} \sum\limits_{n {}={} 1}^N\,\solA{n}{\mu}(x)
\eea
where $\solA{n}{\mu} $ is the solution with one singularity
(not to be confused with $\A{n}{\mu}$ (\ref{Def:An})).
\par
If $\solA{n}{\mu}$ correspond to the solutions of type (\ref{Sol:Move})
then
\be
\solA{n}{\mu} {}={} \sL{n}{0}{.\mu}(x^0) \,\Ssolfi{n}(\rn{n})
\label{Sol:Am}
\ee
where $\Ssolfi{n}(\rn{n})$ is zero component of the electromagnetic potential
for the solution with one rest singularity (\ref{Sol:Stat}),
$\rn{n} {}={} \sqrt{\Ssoly{n}{\! i} \Ssolyd{n}{i}}$
\bea
& &\Ssoly{n}{i}  {}={}  \sL{n}{i}{.j} \left[x^j {}-{} \a{n}{j}(x^0) \right]
\quad;\qquad \a{n}{j}(x^0) {}={} \a{n}{j}(0)
 {}+{} \int\limits_0^{x^0} \V{n}{j}(x^{'0}) \;\d x^{'0}
\label{yLx}\\
& &\Longrightarrow\qquad \frac{\p \Ssoly{n}{i}}{\p x^j} {}={} \sL{n}{i}{.j}
\quad;\qquad \frac{\p \Ssoly{n}{i}}{\p x^0} {}={}
\sL{n}{i}{.0} {}+{} \frac{\d \sL{n}{i}{.j}}{\d x^0} \;
\sL{n}{{-1\,j}}{\phantom{-1\,}.p}\,\Ssoly{n}{p}
\label{dyLx}
\eea
$\sL{n}{\mu}{.\nu}$ is the Lorentz transformation matrix
components (\ref{Def:Lorenz})
that include the velocity of $n$-th singularity $\V{n}{j} {}={} \V{n}{j}(x^0)$.
\par
Thus we have the following expression for initial approximation.
\bea
 F_{\mu\nu}  &=&  \sum\limits_{n {}={} 1}^N\, \left[
\left(\sL{n}{0}{.\nu}\,\frac{\p \Ssoly{n}{i}}{\p x^\mu} {}-{}
\sL{n}{0}{.\mu}\,\frac{\p \Ssoly{n}{i}}{\p x^\nu} \right)
\SsolF{n}{i0}  {}+{}
\left(\delta^0_\mu\,\delta^i_\nu  {}-{}  \delta^0_\nu\,\delta^i_\mu \right)
\frac{\d \sL{n}{0}{.i}}{\d x^0}\,\Ssolfi{n}
\right]\qquad\qquad\\
\mbox{where}\qquad
& &
\SsolF{n}{i0} \,{}\equiv{}\,
\SsolE{n}{i} {}={} \frac{\p \Ssolfi{n}}{\p \Ssoly{n}{i}}
\eea
Let us consider a motion of one $k$-th singularity ($n {}={} k$).
Let us designate a field \mbox{connecting with all other} singularities
($n\neq k$) as $\tilde F_{\mu\nu} $. Thus we have
\be
F_{\mu\nu} {}={}
\left(\sL{k}{0}{.\nu}\,\frac{\p \Ssoly{k}{i}}{\p x^\mu} {}-{}
\sL{k}{0}{.\mu}\,\frac{\p \Ssoly{k}{i}}{\p x^\nu}
\right)
\SsolE{k}{i} {}+{}
\left(\delta^0_\mu\,\delta^i_\nu {}-{} \delta^0_\nu\,\delta^i_\mu \right)
\frac{\d \sL{k}{0}{.i}}{\d x^0}\,\Ssolfi{k}  {}+{}
{\tilde F_{\mu\nu}}
\label{Sol:pert0}
\ee
Let us average these functions $F_{\mu\nu}$ near the $k$-th
singular point. It is evident that
\be
\left\langle
\frac{\p \Ssoly{k}{i}}{\p x^\mu}\,
\SsolE{k}{i} \right\rangle_k {}={} 0\quad;\qquad
\left\langle \Ssolfi{k}\!(\rn{k}) \right\rangle_k {}={}
\Ssolfi{k} \! (0) \quad;\qquad
\left\langle\vphantom{\Ssolfi{k}} {\tilde F_{\mu\nu}} \right\rangle_k {}={}
 {\tilde F_{\mu\nu}} (x^0,\,\bfa{k})
\ee
As can be shown (see also \cite{BornInfeld}),
for used here system of designation we have the following equality for
the solution (\ref{Sol:Stat}).
\bea
\Ssolfi{k}\!(0) {}={} -\beta\,\frac{\en{k}}{\sqrt{|\alpha\,\en{k}|}}\qquad
\mbox{where}\qquad
\beta {}\equiv{} \int\limits_0^\infty\frac{\d r}{\sqrt{1 {}+{} r^4}} {}={}
\frac{\left[\Gamma (\frac{1}{4})\right]^2}{4\,\sqrt{\pi}}
\approx 1.8541
\eea
Let us enter the folloving designation for rest mass of the singularity.
\be
\m{k} {}\equiv{} -\en{k}\;\Ssolfi{k}\!(0) \qquad\quad\left[{}={}
\beta\,\frac{\en{k}{}^2}{\sqrt{|\alpha\,\en{k}|}}
\quad\mbox{for the solution (\ref{Sol:Stat})} \right]
\label{Def:mass}
\ee
Then we can write
\be
\left\langle\vphantom{\bfV{n}} F_{\mu\nu}\right\rangle_k {}={}
-\left(\delta^0_\mu\,\delta^i_\nu {}-{} \delta^0_\nu\,\delta^i_\mu \right)
\;\frac{\m{k}}{\en{k}}\;\frac{\d \sL{k}{0}{.i}}{\d x^0}
 {}+{} {\tilde F_{\mu\nu}}(x^0,\,\bfa{k})
\label{Sol:pert1}
\ee
Substituting (\ref{Sol:pert1}) into the condition (\ref {Eq:Move1})
with $n {}={} k$ we obtain the known Lorentz equation for motion
of $k$-th particle-singularity in initial approximation.
\bea
\m{k} \,\frac{\d}{\d x^0} \frac{ \bfV{k} }{ \sqrt{1 {}-{} \bV2{k}} } {}={}
\en{k}\left({\tilde \E} \,{}+{} \bfV{k}
{} \times {}\, {\tilde \B} \right)
\label{Eq:Move3}
\eea
At the singular point we have $\L {}={} 0$ and $f^{\mu\nu} {}={} \infty$
for the solution (\ref{Sol:Stat}), (\ref{Sol:Move}).
But, as we can show, if we take
$\left(\d \sL{k}{0}{.i}/\d x^0\right) {}={} 0$ in (\ref{Sol:pert0}) then
$\L {}={} 0$, $f^{\mu\nu} {}={} \infty$ for filed $F_{\mu\nu}$
(\ref{Sol:pert0}) on a singular surface
near to the point $\x {}={} \bfa{k}$.
The term with derivative on time from $\sL{k}{0}{.i}$ in (\ref{Sol:pert0})
compensates the field ${\tilde F_{\mu\nu}} $ near to the point
$\x {}={} \bfa{k}$ according to (\ref{Eq:Move3}), so we have the
singular point but not surface.
\par
If other singularities (with $n {}\neq{} k$) are sufficiently distant
from $k$-th one then the field ${\tilde \E},\, {\tilde \B}$
satisfies to the linear Maxwell's system of equations with $n {}\neq{} k$
singularities as sources. Thus we have the standard linear electrodynamics
as initial approximation.
\par
The motion equation (\ref{Eq:Move3}) can be also obtained with
somewhat different way. Let us substitute
$A_\mu  {}={}  \solA{k}{\mu}  {}+{}  {\tilde A_\mu}$ in the action (\ref{Action4})
and make the trajectory variation of the \mbox{$k$-th} singularity only.
Thus, using (\ref{Sol:Am}), (\ref{Def:Lorenz}), (\ref{Def:mass}),
 we obtain the following variational principle that well known
as one for motion of a point charge particle in electromagnetic field.
\bea
\en{k}\,\delta\int \limits_{\tn{k}}
\left(\solA{k}{\mu}  {}+{}  {\tilde A_\mu}\right)\,\d x^\mu  &=&
\delta\Biggl(-\m{k}\int \limits_{-\infty}^\infty
\sqrt{1 {}-{} \bV2{k}}\;\d x^0 {}+{}
\en{k}\int \limits_{\tn{k}}{\tilde A_\mu}\,\d x^\mu \Biggl)\qquad
\nonumber\\
 &=&  \delta\int \limits_{\tn{k}}
\left(-\m{k}\,\d s {}+{} \en{k}\,{\tilde A_\mu}\,\d x^\mu\right)  {}={}  0
\label{Action5}
\eea
where $\d s {}={} \sqrt{|\d x_\nu\,\d x^\nu |}$.
\par
The approach that present here may be applied to any form of
regular part of Lagrangian $(\L {}-{} 1)$. In particular we can
take the Lagrangian for linear theory $\alpha^2\,\I$. It is well known
that in the case of the linear electrodynamics we have a singular solution
with $|\Ssolfi{k}\!(0)| {}={} \infty$. Hence, in this case
$\m{k} {}={} \infty$ and
there is not an interaction between single-singular solutions according
to equation (\ref{Eq:Move3}), that full conforms with
superposition property of solutions for linear theory.
 Thus we can consider the linear electrodynamics only as initial
approximation in a nonlinear theory pursuant to above-stated.
\par
Because all observable charges in Nature multiple to
the electron charge, it would appear reasonable that
$\en{k} {}={} \pm|\mbox{the electron charge}|$.
We can make an any value for the mass (\ref{Def:mass}) of the singularity
with help of the model constant $\alpha$.
But we assume that this mass is not equal to the electron mass because
a possible field configuration coincide to electron
should be a more complex then the solution (\ref{Sol:Stat}). Thus we have
the two model constants $\alpha$ and $e {}\equiv{}  |\mbox{the electron charge}|$.
But they are the dimensional constants and, hence,
we can make $\alpha  {}={}  e  {}={}  1$ for suitable dimensional system.
Thus qualitative characteristics of the model does not depend from
sizes of these constants.

\section{Conservation Laws for Energy-Impulse and Angular Momentum}

Here we shall use a cartesian system of coordinates with components of metric
$h_{\mu\nu}$:\\
$-h_{00} {}={} h_{11} {}={} h_{22} {}={} h_{33} {}={} 1$, $h_{\mu\nu}  {}={}  0$ for $\mu\neq\nu$.
\par
Using the field equation (\ref{EulerMy}), the current conservation
(\ref{ConsLaw:Current0}) and the condition (\ref{Eq:GMove}),
we can show that there is the following differential conservation law for
canonical energy-impulse tensor.
\bea
\frac{\p T^\mu_{.\nu}}{\p x^\mu}  {}={}  0 \quad \mbox{where} \qquad
T^\mu_{.\nu}   {}={}  f^{\mu\rho}\,\frac{\p A_\rho}{\p x^\nu}
 {}-{} \frac{1}{\alpha^2}\left(\L {}-{} 1\right)\,h^\mu_{\nu}  {}+{}
4\pi\,\jmath^{\mu}\,A_\nu
\label{Cons:CanTEI1}
\eea
Then we have a full canonical energy-impulse in the following form.
\bea
P^\nu  {}\equiv{} \frac{1}{4\pi}\,
\int\limits_\Omega T^{0\nu}\,(\d x)^3  {}={}
\frac{1}{4\pi}\,\int\limits_\Omega \left[f^{0\rho}\,\frac{\p A_\rho}{\p x_\nu}
 {}+{} \frac{h^{0\nu}}{\alpha^2}\left(\L {}-{} 1\right)\right]\,(\d x)^3
 {}+{} \sum\limits_{n {}={} 1}^N\, \en{n} \, \Au{n}{\nu}
\eea
As we see, the full canonical energy-impulse is divided into two
parts. The first part has a regular density and the second part
has a singular density. This looks like a field and point particles
in the standard electrodynamics.
According to relations (\ref{Sol:Am}), (\ref{Def:Lorenz}), (\ref{Def:mass})
we have an energy-impulse of $n$-th particle-singularity for initial
approximation in the following form.
\bea
{\P{n}}{}^0 {}={} \frac{\m{n}}{\sqrt{1 {}-{} \bV2{n}}}\quad;\qquad
{\P{n}}{}^i {}={} \frac{\m{n}\,\V{n}{i}}{\sqrt{1 {}-{} \bV2{n}}}
\eea
This expressions coincide with energy-impulse of point particle.
\par
With direct verification we can show that there is the following
differential conservation law for angular momentum tensor.
\bea
\frac{\p M^\mu_{.\nu\rho}}{\p x^\mu}  {}={}   0 \quad
\mbox{where}\qquad
M^\mu_{.\nu\rho}  {}\equiv{}  T^\mu_{.\nu}\,x_\rho  {}-{}  T^\mu_{.\rho}\,x_\nu  {}-{}
 f^\mu_{.\nu}\,A_\rho  {}+{}  f^\mu_{.\rho}\,A_\nu
\eea
\par
Using the definition (\ref{Cons:CanTEI1}) we can write the following
differential conservation law for metric energy-impulse tensor.
\bea
\frac{\p \T^\mu_{.\nu}}{\p x^\mu}  {}={}  0 \quad\mbox{ where }\quad
\T^\mu_{.\nu} &\equiv&  T^\mu_{.\nu}   {}-{}
\p_\rho\,\left(f^{\mu\rho}\,A_\nu \right) {}={}
f^{\mu\rho}\,F_{\nu\rho}  {}-{}
\dfrac{1}{\alpha^2}\,\left(\L {}-{} 1\right)\,h^\mu_\nu
\eea
In this case we have the full metric energy-impulse in the following form.
\bea
\fP^\nu  {}\equiv{} \frac{1}{4\pi}\,
\int\limits_\Omega \T^{0\nu}\,(\d x)^3  {}={}
\frac{1}{4\pi}\,\int\limits_\Omega \left[f^{0\rho}\,F^\nu_{.\rho}
 {}-{} \frac{h^{0\nu}}{\alpha^2}\left(\L {}-{} 1\right)\right]\,(\d x)^3
\eea
As we see the full metric energy-impulse is expressed by a regular
density in this case.

\section{Conclusion}

Thus we have presented an initial theory of nonlinear electrodynamics with
singularities. The main obtained relations do not depend on a form of
relations between the tensor components $f^{\mu\nu}$ and $F^{\mu\nu}$.
Hence, the results are sufficiently general.
\par
The existence of the singularities in this theory is connected with
addition the integrals on the trajectories of the singularities
(\ref{Action4}) to the action (\ref{ActionBI})
that provide also the stability of the singular particle-like solutions.
The singularity may be as
positive as negative depending on whether the sign before the $k$-th
trajectory integral is minus ($\en{k} {}={} e > 0$) or plus
($\en{k} {}={} -e < 0$).
\par
Within the framework of the presented theory
the canonical energy-impulse density is naturally divided on
two part: the regular part ("field") and the singular part ("particles").
The singular part of the energy for rest singularity is equal to the
her rest mass that included in Lorentz equation of motion of the singularity
that we have in initial approximation.
However the metric energy-impulse density is regular.
\par
In initial approximation this theory gives the usual linear
electrodynamics with point charged particles.
\par
The obtained results can be used for search of
the possible field configurations that could describe the real particles.

\section*{Acknowledgements}

I am grateful to Professor A.A.~Grib for useful discussion the results
stated in this article.

\end{document}